\documentclass[aip, reprint, nofootinbib]{revtex4-1}
\usepackage{graphicx}
\usepackage{glossaries}
\usepackage{amsmath}
\usepackage{amssymb,amsmath}
\usepackage[usenames,dvipsnames]{color}
\usepackage[margin=.57in]{geometry}
\usepackage{braket}
\usepackage{enumitem}
\usepackage{mathtools}

\usepackage[english]{babel} 

\usepackage{ntheorem}
\theoremseparator{:}
\usepackage{ulem}

\usepackage{filecontents}
\pdfoutput=1

\begin{document}
\normalem

\title{Topological Description of the Solidification of Undercooled Fluids and the Temperature Dependence of the Thermal Conductivity of Crystalline and Glassy Solids Above Approximately 50 K }

\author{Caroline S. Gorham}
\email{caroling@cmu.edu}
\affiliation{Department of Materials Science and Engineering, Carnegie Mellon University, Pittsburgh, PA 15213, USA}

\author{David E. Laughlin}
\email{laughlin@cmu.edu}
\affiliation{Department of Materials Science and Engineering, Carnegie Mellon University, Pittsburgh, PA 15213, USA}

\begin{abstract}
By the adoption of a quaternion orientational order parameter to describe solidification, the topological origins of the thermal transport properties of crystalline and non-crystalline solid states are considered herein. Global orientational order, achieved by spontaneous symmetry breaking, is prevented at finite temperatures for systems that exist in restricted dimensions (Mermin-Wagner theorem). Just as complex ordered systems exist in restricted dimensions in 2D and 1D, owing to the dimensionality of the order parameter, quaternion ordered systems in 4D and 3D exist in restricted dimensions. Just below the melting temperature, misorientational fluctuations in the form of spontaneously generated topological defects prevent the development of the solid state. Such solidifying systems are well-described using O(4) quantum rotor models, and a defect-driven Berezinskii-Kosterlitz-Thouless (BKT) transition is anticipated to separate an undercooled fluid from a crystalline solid state. In restricted dimensions, in addition to orientationally-ordered ground states, orientationally-disordered ground states may be realized by tuning a non-thermal parameter in the relevant O(n) quantum rotor model Hamiltonian; thus, glassy solid states are anticipated to exist as distinct ground states of O(4) quantum rotor models. Within this topological framework for solidification, the finite Kauzmann temperature marks a first-order transition between crystalline and glassy solid states at a ``self-dual critical point" that belongs to O(4) quantum rotor models. This transition is a higher-dimensional analogue to the quantum phase transition that belongs to O(2) Josephson junction arrays (JJAs). Thermal transport properties of crystalline and glassy solid states, above approximately 50 K, are considered alongside electrical transport properties of JJAs across the superconductor-to-superinsulator transition.
\end{abstract}
\maketitle

\section{Introduction}

The origins of the anomalous temperature dependence of the thermal conductivity of non-crystalline materials as compared with crystalline counterparts have remained a matter of great interest over the past century~\cite{kittel_interpretation_1949, cahill_lattice_1988, orbach_phonon_1993}. In crystalline solid states, i.e., those that exhibit the translational periodicity of a lattice, heat conduction is due to the motion of collective elementary excitations known as phonons~\cite{dove_introduction_1993, kittel_introduction_1996}. Owing to the existence of a well-defined Brillouin zone, above approximately 50 K, phonons are scattered predominantly by resistive Umklapp phonon-phonon scattering processes. In this temperature range, the thermal conductivity increases with decreasing temperatures as resistive phonon-phonon interactions become less frequent. On the other hand, phonons cannot exist in non-crystalline solids that lack a well-defined Brillouin zone. The thermal conductivity of non-crystalline solids decreases with decreasing temperatures over the same temperature range~\cite{eucken_uber_1911, kittel_interpretation_1949}, and agrees well with Einstein's model of a random walk of thermal energy between neighboring groups of atoms~\cite{cahill_lattice_1988, cahill_lower_1992}. 

Herein, we approach the thermal conductivity as an emergent transport property that develops as a consequence of the mechanisms of solidification of undercooled atomic fluids. It is well-known that all solidifying atomic liquids in three-dimensions must undercool below the melting temperature $T_M$, in order to develop the necessary thermodynamic driving force (i.e., change in free energy) for the formation of a solid state. Within the context of thermodynamics, the degree of undercooling (prior to crystallization) is influenced by the rate of nucleation which depends on the size of atomic clusters and the surface energy associated with them. This work goes beyond thermodynamics, by approaching a topological framework for undercooling and solidification that is based on the notions of symmetry breaking. In particular, we focus on the influence of the topological structure of undercooled atomic fluids (that results due to atomic clustering) on the solidification process. In contrast to thermodynamics, for which the mechanisms of solidification are not directly connected to the transport properties of the solid state, it is by considering the role of topology that these phenomena may be related.

An original consideration of the topological nature of undercooling was offered by Frank (Ref.~\onlinecite{frank_supercooling_1952}), who suggested that undercooling below the melting temperature was related to an energetic-preference for icosahedral atomic clustering -- which is incompatible with long-range crystallinity. More recent theoretical and topological approaches to undercooling~\cite{nelson_liquids_1983, sethna_relieving_1983, nelson_symmetry_1984}, that build upon Frank's earlier work, have made use of the fact that particles with icosahedral coordination shells \emph{do} tessellate the surface of a sphere in four-dimensions (despite the incompatibility of local icosahedral order with a space-filling arrangement). In approaching a more complete topological description of solidification in three-dimensions this work, builds upon these original concepts and, adopts a quaternion orientational order parameter to characterize undercooled atomic liquids.

Our approach to three-dimensional solidification/melting follows in the footsteps of original models of two-dimensional melting, proposed by Halperin and Nelson~\cite{halperin_theory_1978}. In these models, topological ordering of an equilibrium concentration of disclination topological defects drives the solidification of an orientationally-ordered crystalline solid state at low-temperatures. Just as in the two-dimensional case~\cite{halperin_theory_1978}, we anticipate that three-dimensional crystalline solid states are achieved due to the formation of complementary disclination pairs which can be regarded as isolated dislocations. Similar ideas~\cite{kosterlitz_ordering_1973}, replacing disclinations with vortices (each are closed loop topological defects), have led to a theory of the transition to the low-temperature phase-coherent state in two-dimensional complex ordered systems (superfluids). Ultimately, this topological framework leads us to an understanding of the inverse behavior of the thermal transport properties of crystalline and non-crystalline solid states above approximately 50 K.

Over the past many decades, there has been a rapid development in the understanding of the role of topology and of the effects of the spatial dimension in which an ordered system exists on its transport behavior~\cite{halperin_resistive_1979, herbut_modern_2007, poran_quantum_2017}. For example, most notably, two-dimensional complex ordered systems (superfluids) are prevented from developing conventional long-range ordering (by spontaneous symmetry breaking) at finite temperatures~\cite{mermin_absence_1966} and are considered to exist in ``restricted dimensions.'' This is owing to the possible existence of an abundance of misorientational fluctuations that take the form of spontaneously generated topological point defects; such model systems demonstrate the properties of a topological transition (Berezinskii-Kosterlitz-Thouless~\cite{berezinskii_destruction_1971, kosterlitz_ordering_1973}) towards a phase-coherent ground state. Herein, the notion of restricted dimensions is extended to the quaternion number domain, i.e., the next higher-dimensional algebra to the complex number domain, which is applied to characterize orientational order in undercooled atomic liquids and derivative solid states~\cite{nelson_symmetry_1984}.  Just as conventional orientational order is prevented for complex ordered systems in $\mathbb{R}^2$, due to the existence of a gas of misorientational fluctuations that take the form of spontaneously generated topological point defects, so to is conventional orientational order prevented for quaternion ordered systems that exist in $\mathbb{R}^4$.

Going a step further, for $n-$vector ordered systems (i.e., complex 2-vector, quaternion 4-vector) that exist in restricted dimensions, there is the possibility for the realization of a distinct phase-incoherent low-temperature state that is a mirror image to the phase-coherent low-temperature state. Prototypical examples are charged complex ordered systems that exist in restricted dimensions, which exhibit a  superconductor-to-superinsulator transition~\cite{gantmakher_superconductor-insulator_2010, vinokur_superinsulator_2008, baturina_superinsulatorsuperconductor_2013} where the phase-incoherent superinsulating ground state has infinite resistance and acts as a mirror image of superconductivity. In these cases, the material transport property of electrical resistance~\cite{poran_quantum_2017} that can be controlled by a non-thermal parameter ($g$) that enters the Hamiltonian~\cite{sachdev_quantum_2011}. In analogue to the superconductor-to-superinsulator transition, a non-thermal transition between orientationally-ordered (crystalline) and orientationally-disordered (non-crystalline) ground states is anticipated in four- or three-dimensions. The thermal transport properties of these crystalline and non-crystalline solid states are compared with the electrical transport properties of $O(2)$ Josephson junction arrays in the vicinity of a superconductor-to-superinsulator transition.

     \begin{figure}[b!]
  \centering
\includegraphics[scale=.4]{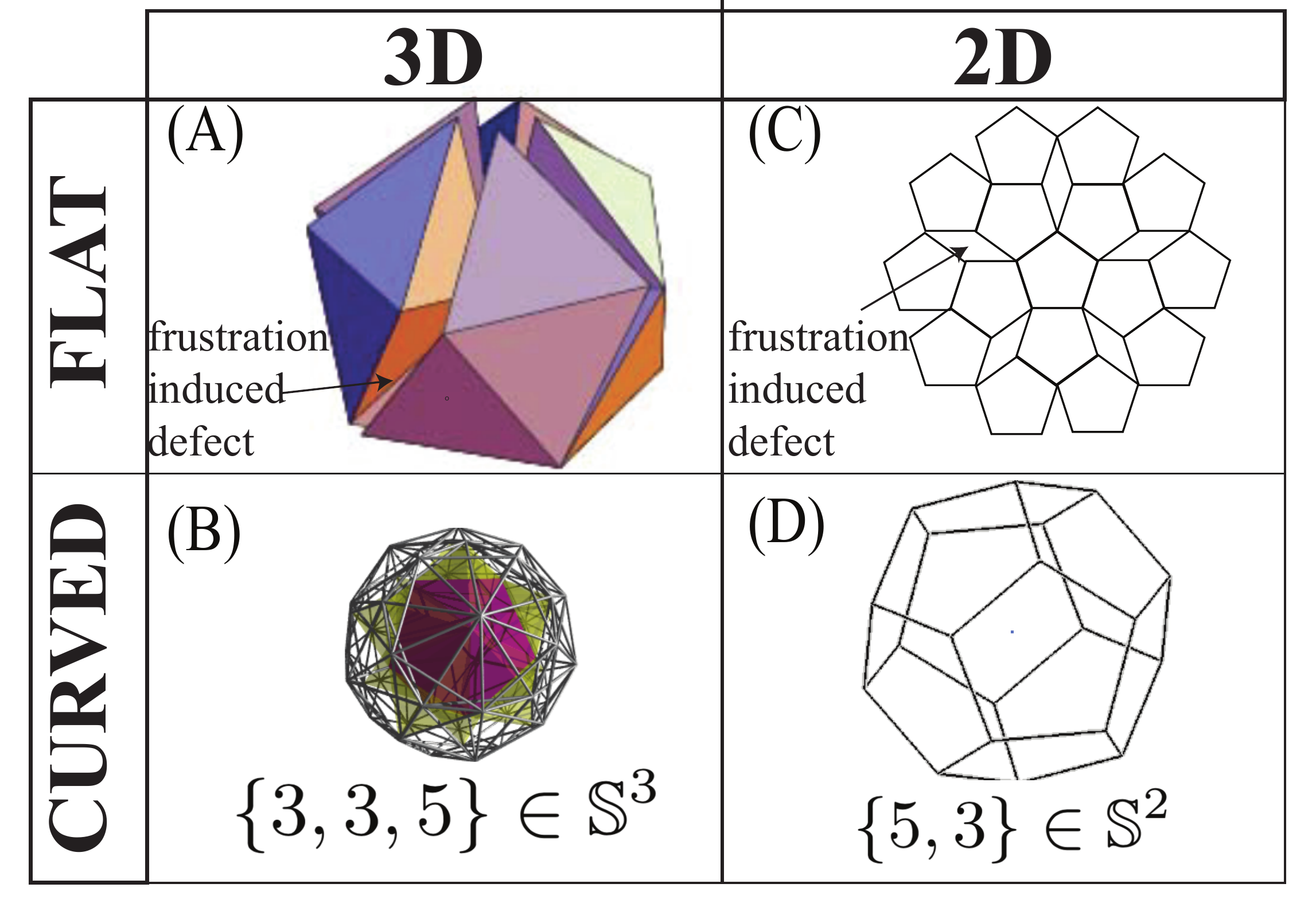}
\caption{(A) and (C) are ``geometrically frustrated'' tilings of three- and two-dimensions, respectively. Just as gaps remain between certain faces when regular tetrahedra are packed into space to form the icosahedral configuration shown here, there is no regular pentagonal tiling of the Euclidean plane. (B) and (D) are regular tessellations of the three-sphere ($\mathbb{S}^3\in \mathbb{R}^4$) and the two-sphere ($\mathbb{S}^2\in \mathbb{R}^3$), by the preferred orientational order in three- or two-dimensions.}
\label{fig:frustrated_plaquette}
\end{figure}

\section{Topological origins of solidification}
\label{sec:topological_solidification}

\subsection{Quaternion orientational order parameter}

At temperatures just below the melting temperature, orientational order in three-dimensional atomic fluids is characterized by the preferred orientational symmetry of atomic clustering~\cite{mermin_homotopy_1978,mermin_topological_1979, nelson_liquids_1983, nelson_symmetry_1984}. This is denoted by the subgroup of three-dimensional rotations $H\in G$, where $G=SO(3)$ is the full orientational symmetry group of the high-temperature liquid. In considering the topological properties of the ordered system, and in order to apply the theorems of homotopy~\cite{mermin_homotopy_1978, mermin_topological_1979}, it is important that the group $G$ be simply connected (i.e., $\pi_1(G)=0$). This larger group, that is simply connected, imbeds the continuous group $G$ (that is not simply connected) and is called the ``universal covering group'' of $G$. The most common example of such a relationship is between the three-dimensional group of proper rotations $SO(3)$ (not simply connected) and the special unitary group of degree two $SU(2)$ of unit quaternion elements~\cite{mermin_homotopy_1978, mermin_topological_1979} (simply connected). The relationship between $SO(3)$ and $SU(2)$ is via a $2-$to$-1$ homomorphism~\cite{mermin_homotopy_1978, mermin_topological_1979}, in which two quaternions correspond to each rotation in three-dimensions.

As a particular example of this $2-$to$-1$ homomorphism, between $SO(3)$ and $SU(2)$, consider the case of preferred icosahedral coordination of atomic clustering about a central atom in three-dimensions $Y\in SO(3)$. As depicted in Figure~\ref{fig:frustrated_plaquette} (A), atoms that express icosahedral coordination are unable to fill three-dimensional space; this phenomenon is known in the literature~\cite{nelson_defects_2002, sadoc_geometrical_2006, griffin_relationship_2017} as ``geometrical frustration.'' However, particles with icosahedral coordination shells \emph{do} tessellate the surface of a sphere in four-dimensions ($\mathbb{S}^3$) forming a four-dimensional Platonic solid known as the $\{3,3,5\}$ polytope (Figure~\ref{fig:frustrated_plaquette} (B)). The 120-vertices of the $\{3,3,5\}$ polytope are the elements of the binary representation~\cite{mermin_homotopy_1978, mermin_topological_1979, stillwell_story_2001} of $Y\in SO(3)$ in $SU(2)$. Similarly, although there is no pentagonal tiling of the plane (Fig.~\ref{fig:frustrated_plaquette} (C)) pentagons do tessellate the surface of a sphere in three-dimensions ($\mathbb{S}^2$) as a dodecahedron (Fig.~\ref{fig:frustrated_plaquette} (D)).    

Ultimately, it is important to note that, as a consequence of the incompatibility of the preferred orientational order of atomic clustering with long-range order (due to short-range constraints~\cite{griffin_relationship_2017} that impose geometrical frustration), there is a mismatch in the Gaussian curvature between flat space and the ideal curved space structure (Figure~\ref{fig:frustrated_plaquette} (B) and (D)). It is this curvature mismatch that forces a finite density of topological defects into the flat space structure (Figure~\ref{fig:frustrated_plaquette} (A) and (C)). These topological defects carry curvature that is proportional to the amount of curvature added to each cell in flat space in order to relieve the geometrical frustration in curved space~\cite{sethna_relieving_1983}.

In order to consider the topological properties of the ordered system, it is important to identify the relevant topological manifold $\mathcal{M}$:
    \begin{equation}
    \mathcal{M}=G/H,
    \label{eqn:m_manifold}
    \end{equation}
 that characterizes the set of degenerate ground states available to the ordered system. In the particular case of solidification, below the melting temperature, atomic clustering breaks the group $G=SO(3)$ to a subgroup $H\in G$. The relevant simply connected group is $G=SU(2)$, such that:
    \begin{equation}
    \mathcal{M}=SU(2)/H'
    \label{eqn:our_manifold}
    \end{equation} 
    where $H'$ is the binary representation of $H$. Thus, the relevant orientational order parameter manifold in the solidification process ($\mathcal{M}$) can be identified with a subgroup of unit quaternion numbers, i.e., $SU(2)$, which gives an algebraic structure to the three-sphere $\mathbb{S}^3$. 
    
Notably, the group of unit quaternions that give an algebra structure to the three-sphere (that resides in $\mathbb{R}^4$) is the next higher-dimensional algebra domain to the unit complex numbers -- that give an algebra system to the unit circle in $\mathbb{R}^2$ (in planar polar coordinates). However, unlike the complex order parameter that applies to superfluids ($\mathbb{S}^1\in \mathbb{R}^2$), the orientational order parameter manifold that applies to solidification is not continuous ($\mathbb{S}^3\in \mathbb{R}^4$) but is instead discrete owing to atomic clustering.

 \subsection{Topological defects and their role in Universality theory of phase transformations}

\subsubsection{Types of available topological defects}

When studying phase transitions, e.g., solidification, it is often important to consider the role played by topological defects that are generated during the symmetry breaking process. Physical insight into the types of available topological defect elements can be gained by considering the topological properties of the relevant ground state manifold that applies to the ordered system (Eqn.~\ref{eqn:m_manifold}). The $i^{th}$ homotopy group $\pi_i(\mathcal{M})$ describes the sets of topological defects given by the possible mappings of an $i-$dimensional sphere around $\mathcal{M}$. Non-trivial homotopy groups contain defect elements that, when drawn on $\mathcal{M}$, cannot be contracted to a point on its surface. 

\begin{figure}[b!]
  \centering
\includegraphics[scale=.7]{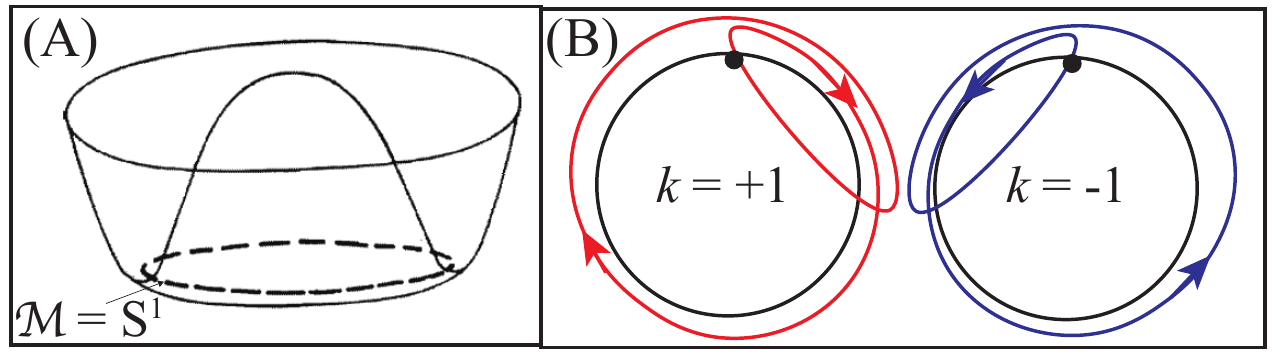}
\caption{(A) The `Mexican hat' potential energy configuration applies to complex ordered systems, below the bulk critical transition temperature. The manifold of degenerate ground states is the locus of points in the complex plane, i.e., $\mathcal{M}=\mathbb{S}^1$.  (B) $\pi_1(\mathbb{S}^1)$ topological defects are available to complex ordered systems, by the specification of $\theta$ on some closed loop. If, on taking a closed loop in space, $\theta$ changes by $2\pi k$ ($k=0,\pm1,\pm2,...$ is the winding number) then a defect core exists within the circuit.}
\label{fig:wrapping_S1}
\end{figure}

Returning to the general case of ordered systems that exhibit a continuous order parameter, the ground state manifold is $\mathcal{M}=S^m$. In this case, only non-trivial homotopy group of defects is~\cite{toulouse_principles_1976}:
\begin{equation}
\pi_m(\mathcal{M})=\mathbb{Z},
\label{eqn:abelian_point}
\end{equation}
where $\mathbb{Z}=0,\pm1,\pm2,...$ is a lattice of integers. Importantly, the $\pi_m(\mathbb{S}^m)$ homotopy groups are necessarily Abelian~\cite{chaikin_principles_2000} because of their identification with $\mathbb{Z}$. Examples of ordered systems that exhibit an $n-$vector order parameter (where $m=n-1$) are particularly interesting to consider. In complex $2-$vector ordered superfluids, $\pi_1(\mathbb{S}^1)$ topological defects (known as vortices) are available~\cite{sethna_statistical_2006} (Fig.~\ref{fig:wrapping_S1}). Similarly, $\pi_3(\mathbb{S}^3)$ topological defects are available to quaternion $4-$vector ordered systems for which the order parameter resides in a four-dimensional vector space.

\begin{figure}[t!]
  \centering
\includegraphics[scale=.7]{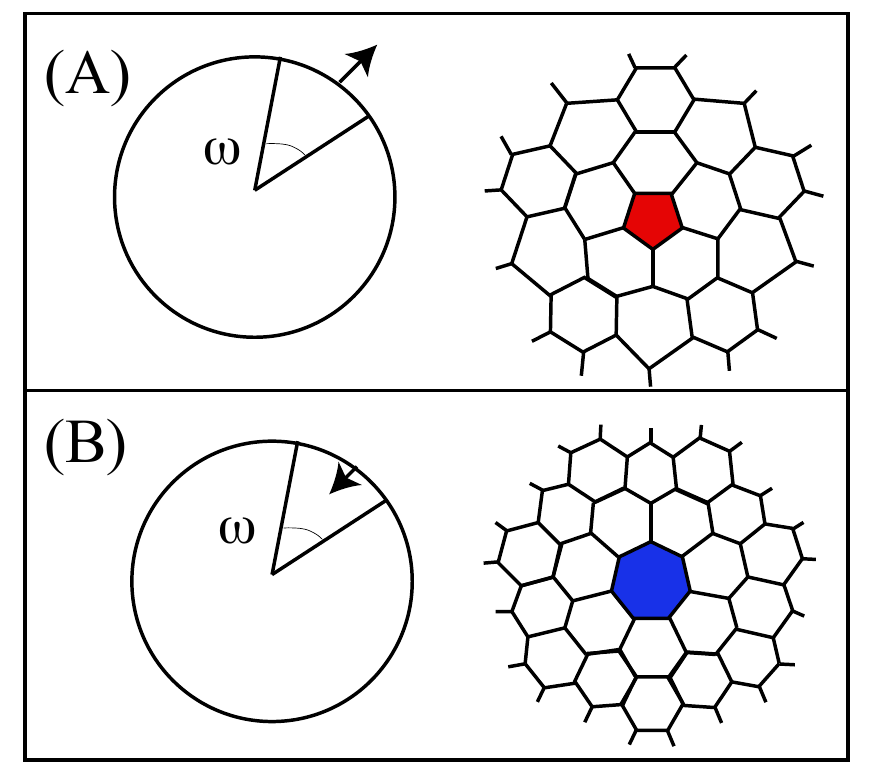}
\caption{Shown are wedge disclinations in a honeycomb net that concentrate (A) positive and (B) negative curvature, respectively. Disclinations, characterized by a Frank vector ($\vec{\omega}$) on taking a measuring circuit around a point in space, are the angular cousin to dislocations which are characterized by a linear Burger's vector. Instead of the Burger's vector, an angular deficit is the topological invariant that attributes a topological strength to a wedge disclination  and imparts curvature to it.}
\label{fig:disclination}
\end{figure}

In real solidifying atomic systems, it is owing to the discrete nature of atomic clustering that the actual ground state manifold is not continuous (i.e., $\mathcal{M}=\mathbb{S}^3$) but instead has points identified on its surface: $\mathcal{M}=\mathbb{S}^3/H'$ where $H'$ is the lift~\cite{mermin_homotopy_1978, mermin_topological_1979} of $H\in G$ into $SU(2)$. This leads to the possible existence of topologically stable defect elements that belong to the fundamental homotopy group, i.e., that are identified by closed loops. These are wedge disclinations, that can be introduced by a standard Volterra process~\cite{hull_introduction_2001}. The Volterra process introduces an angular deficit (Frank vector) that can be measured as a topological invariant on taking a closed loop (measuring circuit) in the sample.  This angular deficit concentrates curvature at the core of the wedge disclination; examples of two-dimensional wedge disclinations in a two-dimensional hexagonal net, that concentrate positive and negative curvature, are shown in Figure~\ref{fig:disclination}.

\subsubsection{Dimensionality of available topological defects}
\label{sec:dimensionality_defects}

\begin{figure}[b!]
  \centering
\includegraphics[scale=.6]{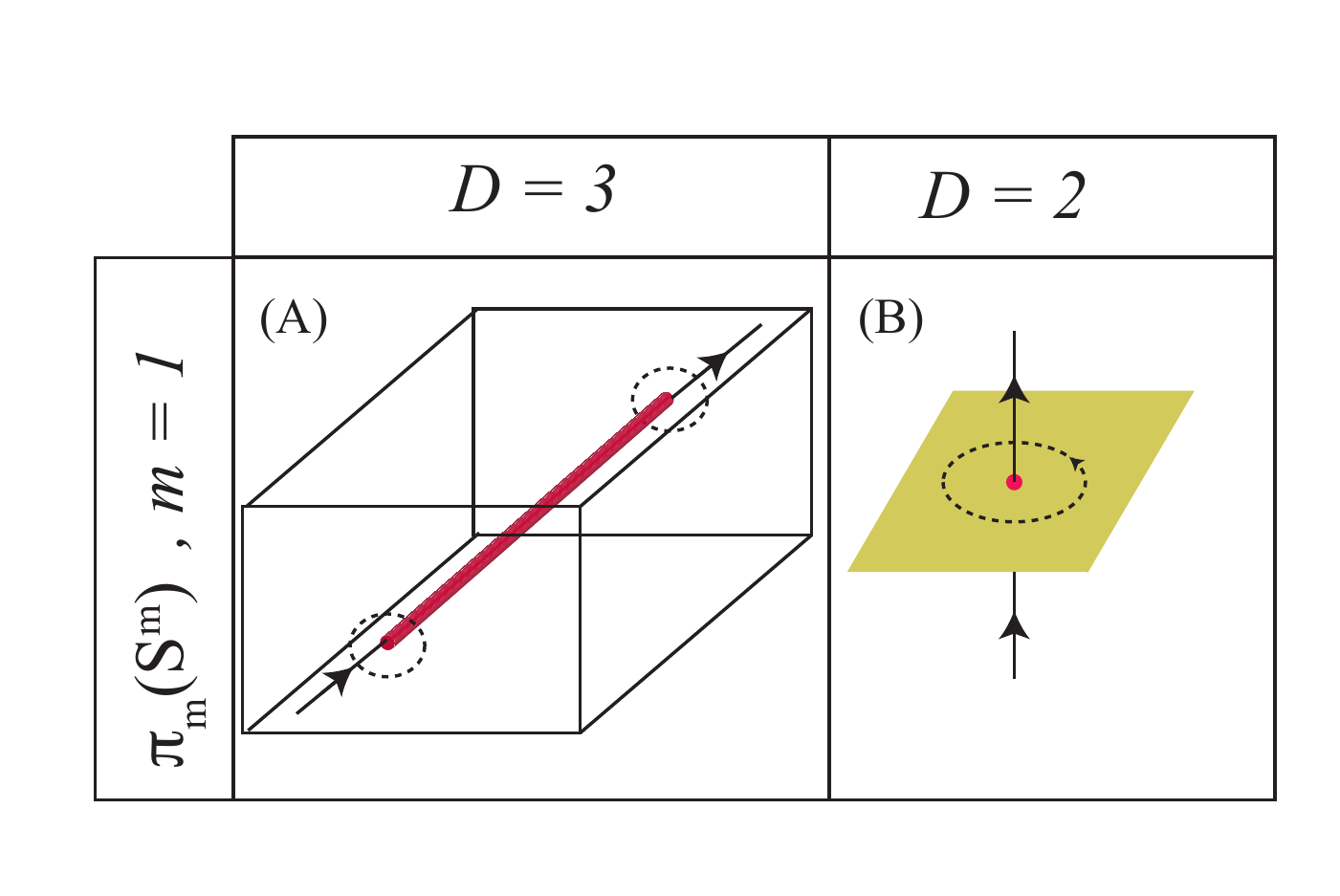}
\caption{Vortex topological defects, that belong to the homotopy group $\pi_1(\mathbb{S}^1)=\mathbb{Z}$ are: (A) linear defects in three-dimensions and, (B) point defects in thin-films. }
\label{fig:topological_charge}
\end{figure}

Once the types non-trivial homotopy groups of topological defects have been identified, it is important to consider how their dimensionality is influenced by the spatial dimension in which the ordered system exists ($D$). This is because the dimensionality of the available topological defects plays an important role in the critical properties of ordered systems, in the vicinity of critical points. In general, defect elements that belong to the homotopy group $\pi_i(\mathcal{M})$, for ordered systems that exist in $D$ spatial dimensions, have the dimensionality~\cite{teo_topological_2010}: 
\begin{equation}
d = D-i-1.
\label{eqn:topocharge}
\end{equation}
Equation~\ref{eqn:topocharge} shows that vortex defects ($\pi_1(\mathbb{S}^1)$), available to complex ordered systems (Fig.~\ref{fig:wrapping_S1}), are linear defects in three-dimensions and are point defects in two-dimensions. Similarly, third homotopy group defects are points in four-dimensions. This case is depicted, for general fundamental homotopy group defects, in Figure~\ref{fig:topological_charge}.

For ordered systems in which the available topological defects are points, conventional orientational order is prevented at finite temperatures~\cite{mermin_absence_1966}. This is owing to the existence of a gas of misorientational fluctuations that take the form of spontaneously generated point defects, that are stabilized by configurational entropy at high-temperatures. The spatial dimension in which topological defects exist as points, known as the lower critical restricted dimension ($D_\text{low}$), is the value of the largest spatial dimension in which conventional global orientational order (by spontaneous symmetry breaking) is no longer possible at finite temperatures~\cite{herbut_modern_2007}. 

\begin{figure}[t!]
  \centering
\includegraphics[scale=.9]{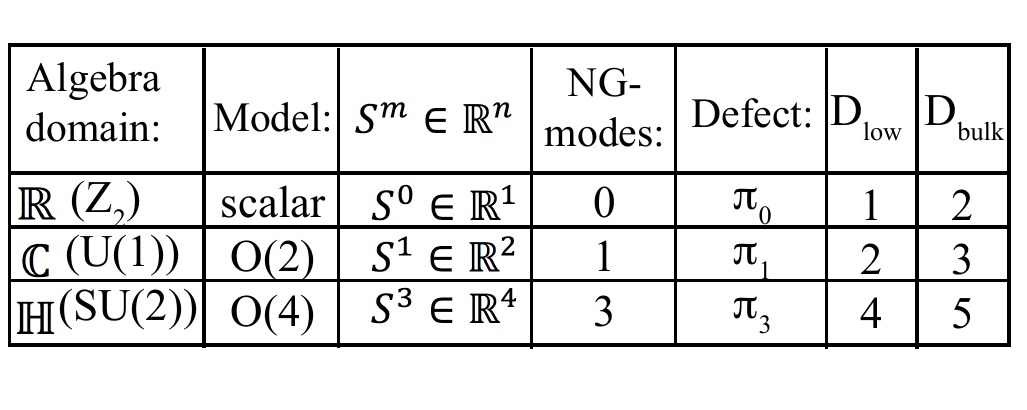}
\caption{Three important algebra domains are: real ($\mathbb{R}$), complex ($\mathbb{C}$), and quaternion ($\mathbb{H}$). The dimension of these number systems are: ($\mathbb{R}$) scalar ($n=1$), ($\mathbb{C}$) two- ($n=2$), and ($\mathbb{H}$) four-dimensional ($n=4$).  For ordered systems characterized by an order parameter that belongs to one of these algebra domains, the relevant order parameter manifold is: $\mathcal{M}=S^m \in \mathbb{R}^n$ where $m=n-1$ (since the amplitude is taken to be constant). These ordered systems have $m$ Goldstone modes and admit $\pi_m$ topological defects~\cite{toulouse_principles_1976}. These topological defects exist as spontaneously generated points~\cite{herbut_modern_2007} in the lower critical dimension $D_\text{low}$, which is the largest dimension in which conventional orientational order by spontaneous symmetry breaking is not possible. All ordered systems characterized by $n$-dimensional order parameters in the dimension $D$ are members of the same ``universality class.''}
\label{fig:phase_trans}
\end{figure}

Identification of the lower critical restricted dimension~\cite{herbut_modern_2007} and of the bulk dimension for ordered systems that are characterized by real ($\mathbb{R}$), complex ($\mathbb{C}$), and quaternion ($\mathbb{H}$) order parameters is made in Figure~\ref{fig:phase_trans}. As the simplest example, for a $\mathbb{Z}_2$ Ising model, the global symmetry is the discrete transformation~\cite{herbut_modern_2007} $\mathbb{Z}_2$ that exchanges ``up'' and ``down'' using the real number domain ($\mathbb{R}$). As a consequence of the symmetry of the $\mathbb{Z}_2$ Ising model, there is an absence of Goldstone modes~\cite{goldstone_broken_1962}. It follows that only $\pi_0$ topological defects exist, which act to prevent conventional global orientational ordering in one-dimensions; therefore, $D_\text{low}=1$ for the Ising model~\cite{herbut_modern_2007}. 

On the other hand, by the \emph{Nambu-Goldstone theorem}~\cite{nambu_quasi-particles_1960, goldstone_broken_1962}, a single Goldstone mode is anticipated in the case of broken $U(1)$ symmetry (by a complex order parameter). It follows that~\cite{toulouse_principles_1976}, for complex ordered systems, vortex ($\pi_1(\mathbb{S}^1)$) are available; fundamental homotopy group elements are topological point defects in two-dimensions, such that $D_\text{low}=2$ for complex ordered systems~\cite{herbut_modern_2007}. Similarly, by Goldstone's theorem, three Goldstone modes are anticipated in the case of broken $SU(2)$ symmetry (by a quaternion order parameter) such that third homotopy group defects ($\pi_3(\mathbb{S}^3)$) are available. Third homotopy group elements are topological points defects in four-dimensions, such that $D_\text{low}=4$ for quaternion ordered systems.

In contrast to ordered systems that exist in restricted dimensions, in bulk ordered systems that exist in a dimension larger than $D_\text{low}$, conventional orientational order develops by spontaneous symmetry breaking. This is possible owing to the higher-dimensional nature of $\pi_m(\mathbb{S}^m)$ topological defects in bulk spatial dimensions (as compared with points in $D_\text{low}$). In bulk dimensions, owing to the fact that the free energy cost to introduce higher-dimensional topological defects is too high to permit them in the absence of applied external fields~\cite{halperin_resistive_1979}, the existence of a finite density of $\pi_m(\mathbb{S}^m)$ topological defects at finite temperatures does not prevent spontaneous symmetry breaking.

\subsubsection{Universality classes and $O(n)$ quantum rotor models}

The effects of the spatial dimension ($D$) in which an ordered system exists on its long-range order (LRO) properties, as discussed above (Section~\ref{sec:dimensionality_defects}), are anticipated by the Universality hypothesis~\cite{herbut_modern_2007}. The Universality hypothesis predicts that the critical behavior of all ordered systems that are characterized by $n$-dimensional order parameters and exist in $D$ dimensions, i.e., that belong to the same ($D$, $n$) ``universality class,'' will be the same in the vicinity of critical points~\cite{herbut_modern_2007}. While phase transitions in bulk dimensions (i.e., $D> D_\text{low}$) occur as a result of spontaneous symmetry breaking (i.e., conventional), phase transitions in restricted dimensions (i.e., $D\leq  D_\text{low}$) are necessarily topological (i.e., BKT). In general, below a bulk critical transition temperature, $n-$vector ordered systems that exist in a restricted dimension ($D$) are well-described mathematically using $O(n)$ quantum rotor models~\cite{sachdev_quantum_2011, herbut_modern_2007} in $D-$dimensions.

Each $O(n)$ quantum rotor, represents the internal orientational degrees of freedom for the low-energy states of a small number of condensed particles~\cite{sachdev_quantum_2011} and, is weakly-linked to its nearest-neighbors. For $n-$vector ordered systems that exist in restricted dimensions, the question of understanding the possible low-temperature ordered states that are able to exist remains an important problem in condensed matter physics. The simplest and most well-studied of such systems that exist in restricted dimensions are charged complex ordered systems with $O(2)$ symmetry~\cite{jose_40_2013}, i.e., Josephson junction arrays. This example generalizes to systems with $O(n)$ symmetry~\cite{herbut_modern_2007}. 

Critically, $O(n)$ quantum rotor model Hamiltonians consist of both potential and kinetic energy terms~\cite{sachdev_quantum_2011} that cannot be minimized simultaneously. The potential energy term favors a ground state of perfectly aligned order parameters~\cite{sachdev_quantum_2011}, and therefore establishes global orientational order at low-temperatures. In contrast, the kinetic energy term favors the localization of condensed particles and a maximally orientational disordered low-temperature state. It is the competition between these two energies that leads to a diverse spectrum of possible ground states for $n-$vector ordered systems that exist in restricted dimensions. This spectrum of possible ground states will be discussed, for complex and quaternion ordered systems, in the sections that follow (Sections~\ref{sec:defect_BKT}, \ref{sec:geometrical_frustration} and \ref{sec:glass_BKT}).

\subsection{$O(n)$ quantum rotor model: defect-driven Berezinskii-Kosterlitz-Thouless (BKT) transformation}
\label{sec:defect_BKT}

Just below a bulk critical transition temperature, in the range of dominant potential energy of the relevant $O(n)$ quantum rotor model Hamiltonian,  a gas of misorientational fluctuations develops that takes the form of a plasma of topological defects whose mobility prevents the development of phase-coherence (orientational order) at finite temperatures. In these systems, despite the prevention of conventional long-range order at finite temperatures (Mermin-Wagner theorem~\cite{mermin_absence_1966}), an orientationally-ordered ground state develops by the minimization of the potential energy term via a defect-driven topological ordering transition, of the Berezinskii-Kosterlitz-Thouless (BKT) type~\cite{berezinskii_destruction_1971, kosterlitz_ordering_1973}. Such defect-driven topological ordering transitions have been most well-studied in classical 2D $O(2)$ Josephson junction arrays~\cite{fazio_charge-vortex_1992} of complex ordered systems.

Despite the prevention of conventional long-range order at non-zero temperatures~\cite{mermin_absence_1966}, the ground state that is obtained by the minimization of the potential energy term of $O(n)$ quantum rotor models is one of perfectly aligned order parameters~\cite{sachdev_quantum_2011}. The relevant potential energy term is developed by placing a single $O(n)$ quantum rotor on the sites $i$ of a $D$ dimensional lattice for which nearest-neighbors interact across weak-links:
\begin{equation}
\hat{V} = - J \sum_{\langle \text{ij} \rangle } \hat{n}_\text{i} \cdot \hat{n}_\text{j},
\label{eqn:general_potentialE}
\end{equation} 
where $\hat{n}_\text{i}$ is the orientation of the  rotor located at site $i$ (e.g., $\mathbb{S}^1\in \mathbb{R}^2$ or $\mathbb{S}^3/H'\in \mathbb{R}^4$),  $J$ is the interaction energy, and the sum is taken over all nearest-neighbor interactions $\langle \text{ij} \rangle$. 

Orientationally-ordered low-temperature states that exist in restricted dimensions, that have been obtained via a defect-driven BKT-type topological transition, are inherently `quasi' long-range ordered at finite temperatures in so far as excitations from the ground state are in the form of bound pairs of topological defects~\cite{herbut_modern_2007, jakubczyk_thermodynamics_2016}. Topological arguments suggest that, because these point defects are necessarily Abelian~\cite{chaikin_principles_2000}, it is possible to anticipate the formation of ``sum-0'' bound pairs. By the minimization of potential energy, below a critical finite temperature, these low-energy sum-0 bound pairs become energetically favorable thereby allowing for the realization of an orientationally-ordered low-temperature state.

It has recently been suggested that~\cite{gorham_su2_2018}, as a higher-dimensional analogue to $O(2)$ quantum rotor models, the formation of low-energy bound states of third homotopy group point defects allows for the development of orientationally-ordered low-temperature states of four-dimensional quaternion ordered systems (that are well-modeled using $O(4)$ quantum rotor models).  Although third homotopy group defects have been previously identified as a non-trivial homotopy group of broken $SO(3)$ symmetry (Ref.~\onlinecite{rivier_gauge_1990}), their pivotal role in solidification has yet to be worked out and is pursued herein.

\subsubsection{Binding of $\pi_m(S^m)$ topological defects}
        
For ordered systems that exist in restricted dimensions, defect-driven topological transitions occur as a consequence of the existence of a plasma of spontaneously generated $\pi_m(\mathbb{S}^m)$ defects. $\pi_m(\mathbb{S}^m)$ topological defects are necessarily Abelian~\cite{chaikin_principles_2000}, such that the rule for their combination is the addition of their topological strengths (e.g., winding number). Thus, a path $\Gamma$ that encloses two single $\pi_m(\mathbb{S}^m)$ defects with topological strengths $k_1$ and $k_2$ is homotopic to the two paths enclosing the point defects individually~\cite{chaikin_principles_2000}. Such bound pairs of topological defects whose strengths sum to zero, known as ``sum-0'' pairs, are topologically equivalent to the uniform state; these paired configurations are energetically favored over isolated topological defects at low-temperatures.

\paragraph{Complex $2-$vector ordered systems: }

\begin{figure}[t!]
  \centering
\includegraphics[scale=.32]{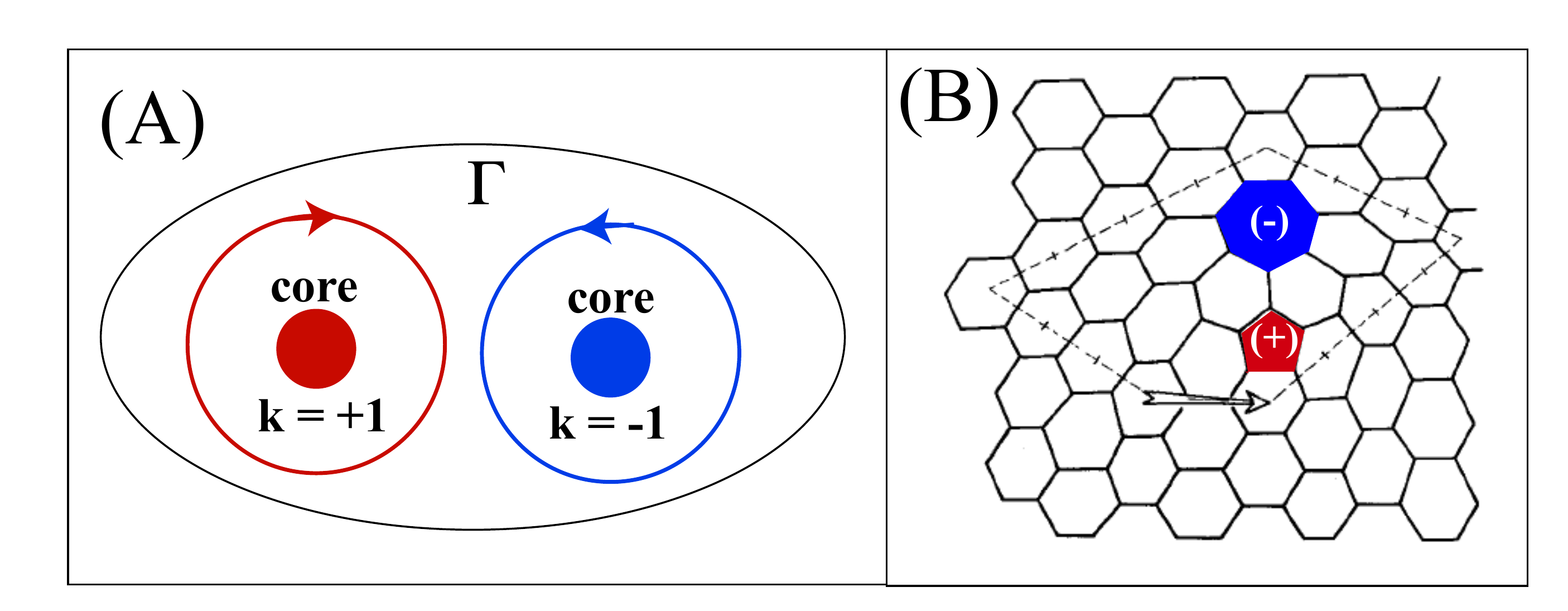}
\caption{(A) A loop $\Gamma$ that surrounds a pair of point defect cores whose topological invariants sum to zero is topologically equivalent to the uniform state, because $\pi_m(S^m)$ defects are Abelian. (B) A dipole of positive and negative wedge disclinations, and the Burger's circuit and Burger's vector used to define the equivalent edge dislocation [Fig. after Refs.~\onlinecite{sadoc_orientational_1985} and~\onlinecite{yazyev_polycrystalline_2014}]. }
\label{fig:core_BKT}
\end{figure}

A most notable example of topological ordering in restricted dimensions is the case of two-dimensional classical Josephson junction arrays (superfluids), for which conventional long-range order is prevented at finite temperatures (Mermin-Wagner theorem). Such systems are physical realizations of $XY-$models, that demonstrate the properties of a Berezinskii-Kosterlitz-Thouless~\cite{kosterlitz_ordering_1973} (BKT) topological ordering transition towards a low-temperature phase-coherent state. This defect-driven BKT transition is driven by the formation of low-energy bound pairs of topological point defects below a finite critical temperature; a cartoon depicting the formation of a sum-0 pair of point defects in two-dimensions (i.e., $\pi_1(\mathbb{S}^1)=\mathbb{Z}$) is shown in Fig.~\ref{fig:core_BKT} (A). To reiterate, it is the formation of these low-energy pairs of $\pi_1(\mathbb{S}^1)$ defects (via the prototypical two-dimensional defect-driven BKT transition~\cite{kosterlitz_ordering_1973}) that allows for the existence of a phase-coherent low-temperature state.

In order to determine the critical transition temperature, at which sum-0 bound pairs become energetically favorable, one must consider the minimization of the potential energy (Eqn.~\ref{eqn:general_potentialE}) that arises due to coupling between nearest-neighbor $O(2)$ rotors that represent complex order parameters. The orientation of each rotor $O(2)$ rotor located at site $i$ is determined by its scalar phase angle parameter $\theta_\text{i} \in [0,2\pi]$: $\hat{n}_\text{i} = (\cos\theta_\text{i}, \sin\theta_\text{i}).$ Thus, the potential energy term of $O(2)$ quantum rotor model becomes:
\begin{equation}
\hat{V} = -J\sum_{\langle \text{ij} \rangle} \cos(\theta_\text{i} - \theta_\text{j}).
\label{eqn:xy_potential}
\end{equation}
At temperatures just below the bulk critical transition temperature, on the formation of an $O(2)$ Josephson junction array, misorientational fluctuations in the scalar phase angles  ($\theta_\text{i}$) throughout the system prevent global phase-coherency. In two-dimensional systems, these misorientational fluctuations take the form of topologically stable point defects (vortices) that must undergo a topological ordering event in order to allow for the existence of a phase-coherent state at finite temperatures~\cite{kosterlitz_ordering_1973}. In the phase-coherent low-temperature state, ordered regions within the array display macroscopic complex order parameters and their the separate phases will be locked together by strong Josephson coupling~\cite{tinkham_physical_1989}.

\paragraph{Quaternion $4-$vector ordered systems: }

Four-dimensional $O(4)$ quaternion ordered systems are a higher-dimensional analogue to two-dimensional $O(2)$ Josephson junction arrays. Just below a bulk critical transition temperature, in the range of dominant potential energy (Eqn.~\ref{eqn:general_potentialE}), conventional global orientational order is prevented due to the existence of a gas of misorientational fluctuations that takes the form of third homotopy group topological point defect elements, i.e., $\pi_3(\mathbb{S}^3)$. In analogue to complex ordered systems that exist in restricted dimensions, is anticipated that, a defect-driven BKT topological ordering transition of these topological point defects allows for the existence of an orientationally-ordered low-temperature state~\cite{gorham_su2_2018}. Just as in the case of complex ordered systems, the formation of low-energy ``sum-0'' pairs of $\pi_m(\mathbb{S}^m)$ point defects in four-dimensional classical $O(4)$ rotor models can be anticipated because $\pi_m(\mathbb{S}^m)$ homotopy groups are necessarily Abelian~\cite{gorham_su2_2018}.

Below the bulk critical transition temperature, the orientation of each $O(4)$ rotor is determined by three scalar phase angle parameters ($\theta \in [0,\pi], \theta_1 \in [0,\pi], \theta_2 \in [0,2\pi]$):   $\hat{n}_i= (\cos\theta_\text{i}, \sin\theta_\text{i}\cos\theta_\text{1,i},  \sin\theta_\text{i}\sin\theta_\text{1,i}\cos\theta_\text{2,i},  \sin\theta_\text{i}\sin\theta_\text{1,i}\sin\theta_\text{2,i}  ).$ The potential energy (Eqn.~\ref{eqn:general_potentialE}) of $O(4)$ quantum rotor model becomes:
\begin{multline}
\hat{V} = -J\sum_{\langle \text{ij} \rangle}  (\cos\theta_\text{i}\cos\theta_\text{j} + \sin\theta_\text{i}\sin\theta_\text{j}\times \\ (\cos\theta_\text{1,i}\cos\theta_\text{1,j} + \sin\theta_\text{1,i}\sin\theta_\text{1,j}[\cos(\cos\theta_\text{2,i}- \cos\theta_\text{2,j})]   )).
\label{eqn:xyzw_potential}
\end{multline}
As the temperature of this system is lowered towards 0 K, this potential energy function is minimized by a ground state of perfectly aligned order parameters. This is a direct higher-dimensional analogue to the two-dimensional $XY$ model (Eqn.~\ref{eqn:xy_potential}), and an alternative form of a phase transition towards the orientationally-ordered low-temperature state should occur that is topological and defect-driven~\cite{gorham_su2_2018}.

This anticipated transition, towards an orientationally-ordered low-temperature state of a four-dimensional quaternion ordered system, was recently studied computationally by the authors (Ref.~\onlinecite{gorham_su2_2018}) using Monte-Carlo simulations. The computational results display characteristic behavior of a topological ordering phase transition, towards a ground state of perfectly aligned $O(4)$ rotors; the authors attributed this~\cite{gorham_su2_2018} to the binding of third homotopy group point defects into low-energy pairs, below a critical finite temperature.

\subsubsection{Binding of $\pi_1(\mathcal{M})$ topological defects}

In real solidifying atomic systems, owing to the discrete symmetry of atomic clustering, wedge disclination topological defects (fundamental homotopy group, Fig.~\ref{fig:disclination}) are also present in the gas of misorientational fluctuations that develops below the melting temperature. Because disclination topological defects belong to the fundamental homotopy group, by Eqn.~\ref{eqn:topocharge}, they are: point-like in 2D, linear in 3D and planar in 4D. In three-dimensions, for example, wedge disclination defects are created by connecting measuring circles surrounding points in a plane~\cite{alexander_colloquium:_2012} along a line. 

Just like third homotopy group defects, wedge disclinations are topologically stable (when drawn on $\mathcal{M}=\mathbb{S}^3/H'$) and therefore must undergo a topological ordering process in order to allow for the realization of an orientationally-ordered crystalline solid low-temperature state. Pairs of low-energy complementary wedge disclinations (Fig.~\ref{fig:core_BKT} (B)), that are favored energetically  (over isolated wedge disclinations) below the finite temperature that marks crystallization, are considered to be edge dislocations~\cite{landau_theory_1986, chaikin_principles_2000, pretko_fracton-elasticity_2018, yazyev_polycrystalline_2014} that belong to the fundamental homotopy group of the order parameter space of the low-temperature crystalline solid state. 

In condensed matter physics, the topological properties of the order parameter manifold $\mathcal{M}$ play a critical role in the emergent transport properties~\cite{teo_topological_2017, griffin_relationship_2017} of the ordered system. Here, the most important example to consider when developing topological arguments for the binding of wedge disclinations (that carry curvature) into edge dislocations (that carry no curvature) is how the topological properties of the ordered system changes during the crystallization process (i.e., undercooled atomic liquids ($\mathbb{S}^3$) to the crystalline solid low-temperature state ($\mathbb{T}^3$)). 

Topology is concerned with the properties of a manifold ($\mathcal{M}$) that are preserved under continuous deformations. A particularly important topological invariant property is the genus of $\mathcal{M}$, that measures the number of ways that you can cut slices of $\mathcal{M}$ without it falling apart (often described as the number of holes in $\mathcal{M}$). Topologically, all manifolds $\mathcal{M}$ are classified according to their genus ($g$) which has a strong relationship to the curvature of the surface. 

\begin{figure}[t]
  \centering
\includegraphics[scale=.9]{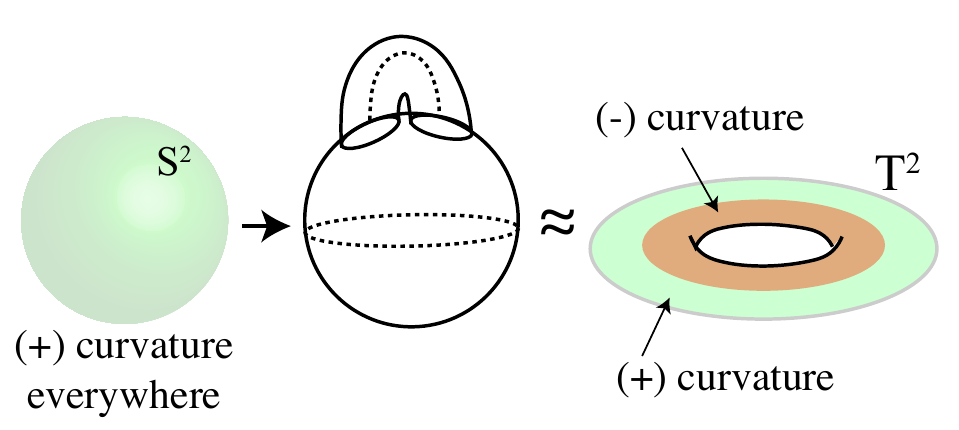}
\caption{An $m-$dimensional sphere ($\mathbb{S}^m$) can be transformed into an $m-$dimensional torus ($\mathbb{T}^m$) by ``adding a handle'' to the surface, i.e., the genus of $\mathbb{T}^m$ is one while the genus of $\mathbb{S}^m$ is zero. This is shown here in two-dimensions. The surface without handles ($\mathbb{S}^m$) has positive integral Gaussian curvature, while the total curvature of the surface with a single handle ($\mathbb{T}^m$) is zero no matter how the surface is deformed.}
\label{fig:euler_characteristic}
\end{figure}

The total curvature of the surface can change only if the topology of the surface changes, for instance, by adding a handle to the surface in order to change its genus (Fig.~\ref{fig:euler_characteristic} (A)). The relationship between topology and curvature is most easily seen by considering two-dimensional orientable surfaces (e.g., $\mathbb{S}^2$ and $\mathbb{T}^2$), for which the relationship between the genus and curvature of a surface ($\mathcal{M}$) follows the Gauss-Bonnet theorem:
\begin{equation}
\int_\mathcal{M}K dA = 2\pi \chi
\label{eqn:gauss_bonnet}
\end{equation}
where $\chi=2(1-g)$ is the Euler characteristic of $\mathcal{M}$, and $K$ is its Gaussian curvature.

Consider the two-dimensional sphere ($\mathbb{S}^2$), for example, for which the Gaussian curvature is constant everywhere. By Eqn.~\ref{eqn:gauss_bonnet},  the integral of the Gaussian curvature is just the area times the constant positive curvature (i.e., $4\pi R^2\cdot 1/R^2 = 4\pi$) such that the Euler characteristic of a two-dimensional sphere is $\chi = 2$. By adding a single handle to an $n-$dimensional sphere, one obtains an $n-$dimensional torus (see Fig.~\ref{fig:euler_characteristic} (A)) whose Euler characteristic is zero (i.e., $\chi=0$). By Eqn.~\ref{eqn:gauss_bonnet}, such a surface (that has the topology of a sphere with a single handle) has zero integral curvature~\cite{hatcher_algebraic_2002}. This is true no matter how the surface is deformed, such e.g., an $n-$dimensional torus and an $n-$dimensional coffee mug (with a single handle) are homeomorphic.

 \begin{figure}[b]
  \centering
\includegraphics[scale=.8]{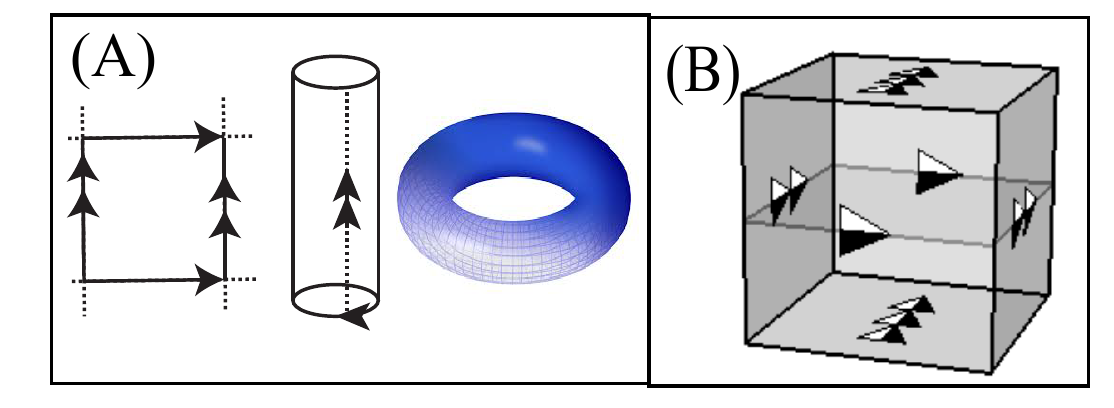}
\caption{(A) A generic 2D Brillouin zone, with maximum wave-vectors $k_x$ and $k_y$. Owing to periodic boundary conditions that apply to the first Brillouin zone, it has the topology of a two-dimensional torus $\mathbb{T}^2$.  (B) A generic 3D Brillouin zone, with periodic boundary conditions.}
\label{fig:torus_coffeecup}
\end{figure}

           \begin{figure}[t!]
  \centering
\includegraphics[scale=.65]{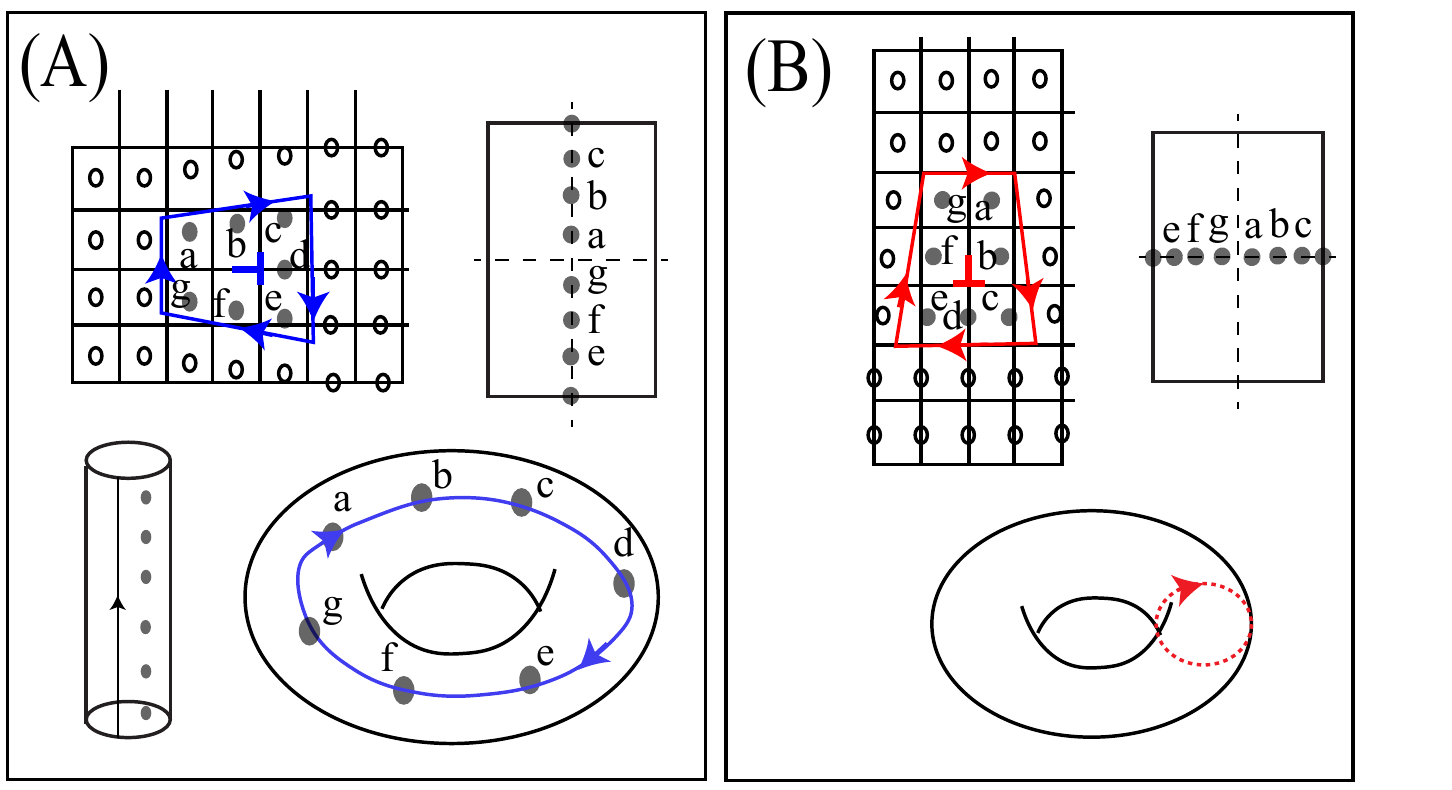}
\caption{Dislocation fundamental homotopy group defects are permitted in crystalline solids. As a closed loop is traversed around a dislocation, the positions of atoms drift with respect to their ideal lattice positions. This corresponds to a loop around the order parameter space. Rearranging the atoms slightly deforms the loop, but does not change the number of times it wraps around the torus. (A) A loop around a dislocation that corresponds to an extra row of atoms corresponds to a path that passes around the hole of the torus. (B) A path through the hole of the torus corresponds to an extra column of atoms. [Figure after Ref.~\onlinecite{sethna_statistical_2006}]}
\label{fig:torus_disclination}
\end{figure}

On crystallization, a ground state of ideal lattice points develops that is characterized by a well-defined Brillouin zone (primitive cell in reciprocal space) in $D$-dimensions. The Brillouin zone exhibits periodic boundary conditions (Born-von Karman~\cite{dove_introduction_1993}), and acts to define the momentum order parameter space of the crystalline lattice. For crystalline solid states, owing to these periodic boundary conditions (Born-von Karman), the topology of the order parameter space is a $D-$dimensional torus. A two-dimensional example is shown in Fig.~\ref{fig:torus_coffeecup} (A), for which a two-dimensional torus is obtained by gluing opposite edges of a 2D Brillouin zone together according to boundary conditions.

The change in topology of the order parameter space during the crystallization process (i.e., the development of a ground state of translational order from an undercooled atomic liquid) may be viewed as akin to the adding of a handle onto the orientational order parameter space of the undercooled atomic liquid. The change in topology of the order parameter space at the crystallization transition in three-dimensions follows directly from the  two-dimensional example depicted in Fig.~\ref{fig:euler_characteristic} (A). The order parameter space of a three-dimensional crystalline solid state is a three-dimensional Brillouin zone with periodic boundary conditions (Fig.~\ref{fig:torus_coffeecup} (B)). A three-dimensional torus ($\mathbb{T}^3$) is obtained when the opposite faces are glued together according to the boundary conditions.

Owing to the development of translational periodicity (i.e., a well-defined Brillouin zone), the thermal entropy due to the displacement of atoms away from their ideal lattice positions at finite temperatures is characterized by the density of collective excitations (known as phonons). In real three-dimension crystalline solids, phonons are the three anticipated Nambu-Goldstone modes of broken quaternion symmetry that are classified as 1 longitudinal mode (compression waves) and 2 transverse modes (shear waves) per condensed atom. The spectrum of phonons describe the energy cost to perform certain allowable atomic displacements from the crystalline ground state, owing to the rigidity of the ordered system. Atomic displacements that give rise to phonon collective excitations can be drawn on the surface of $\mathcal{M}$, and may always be contracted to a point on its surface; in this way, phonon excitations are unique from topological defects, which are topologically stable on the surface $\mathcal{M}$.

In addition to elementary excitations in crystalline solid states (phonons), \emph{topological defects} in the translational order parameter field (i.e., dislocations) may also be present at finite temperatures. As noted earlier, the order parameter space of a crystalline solid, is an $n-$dimensional torus shape ($\mathbb{T}^n$). Topologically, $n-$dimensional tori are homeomorphic to the Cartesian product of $n$ circles: $\underbrace{\mathbb{S}^1 \times ... \times \mathbb{S}^1}_{\text{n}}$. It follows that, the fundamental homotopy group of $\mathbb{T}^n$ is isomorphic to the product of the fundamental homotopy group of $n$ circles~\cite{hatcher_algebraic_2002}. Hence, the fundamental homotopy group in three-dimensional crystalline solids is:
\begin{equation}
\pi_1(\mathbb{T}^3) = \mathbb{Z} \times \mathbb{Z}\times \mathbb{Z},
\end{equation}
where $\mathbb{Z}$ represents the set of all arbitrary integers. That is, a dislocation defect in three-dimensions is labeled by a set of three integers ($b_u$, $b_v$, $b_w$) that identify the magnitude of the Burger's vector in the $x-$, $y-$ and $z-$ directions.

Two dimensional examples of dislocations are shown in Fig.~\ref{fig:torus_disclination}, for which the field of atomic displacements that is introduced in the vicinity of a dislocation defect core corresponds to a loop around the hole of a two-dimensional torus. Rearranging the atoms slightly deforms the loop but does not change the number of times it wraps around the hole; this is why a dislocation is a topological defect. Importantly, unlike $\pi_1$ disclinations, $\pi_1$ dislocations carry no curvature. Topologically, this is a fundamental consequence of the fact that the tori, i.e., a sphere with a single handle, upon which dislocation closed-loop defects are defined have zero integral Gaussian curvature. Because the $n-$dimensional torus has zero integral Gaussian curvature, any closed-loop topological defect that is defined upon its surface is also flat, i.e., it carries no curvature.

Thus, it is anticipated that, at the finite temperature that marks crystallization, pairs of complementary wedge disclination defects (with equal and opposite angular deficit) bind together. These bound pairs of complementary wedge disclinations, that are topologically stable as isolated deects in the undercooled atomic liquid, are topologically equivalent to edge dislocations that are characterized by a Burger's circuit (Fig.~\ref{fig:core_BKT} (B)). Isolated wedge disclinations are only able to persist within the solid state in the event of geometrical frustration~\cite{nelson_symmetry_1984, sadoc_geometrical_2006} (Fig.~\ref{fig:frustrated_plaquette}), as will be discussed more completely in Section~\ref{sec:geometrical_frustration}.

\subsection{Geometrical Frustration and the Major Skeleton Network in Topologically Close-Packed Crystalline Solids}
\label{sec:geometrical_frustration}

    \begin{figure}[t!]
  \centering
\includegraphics[scale=.4]{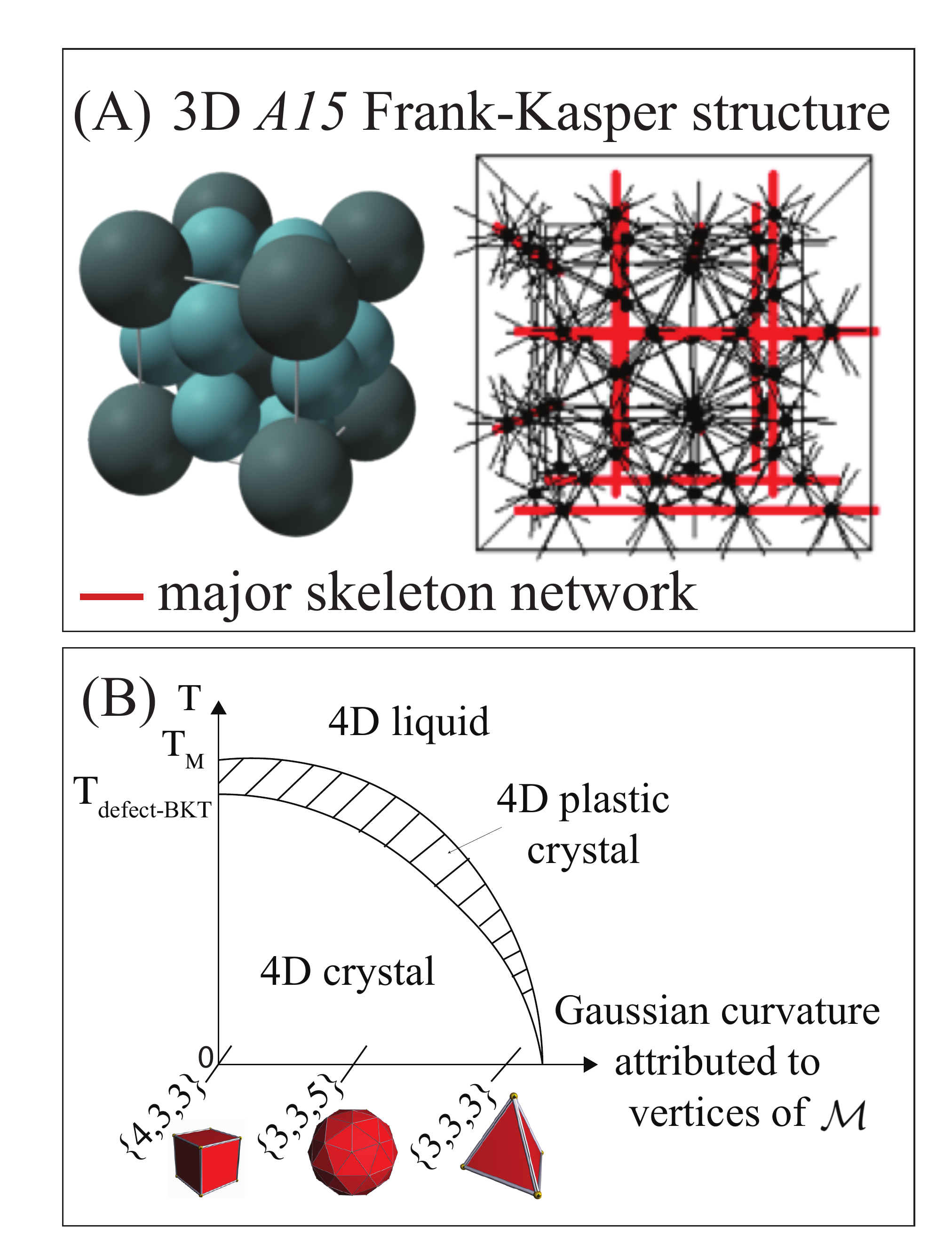}
\caption{ (A) $A15$ structures are one of the most common of the  Frank-Kasper family of TCP phases. Three orthogonal grids of $Z=14$ coordinated atoms, which are $-72^\circ$ disclination lines~\cite{nelson_liquids_1983}, comprise the ordered major skeleton network in $A15$ compounds. [Figure after Ref.~\onlinecite{sadoc_geometry_1990}] (B) Anticipated solidification phase diagram of a 4D plastic crystal, in coordinates $T$ vs. geometrical frustration, in the range of dominant potential energy.}
\label{fig:curving_wedge}
\end{figure}

In solidification, the role of geometrical frustration is to introduce finite kinetic energy effects into the relevant $O(4)$ quantum rotor model Hamiltonian. Thus, in the range of dominant potential energy, geometrical frustration may be viewed as akin to magnetic frustration that can be realized in charged $O(2)$ Josephson junction arrays in the presence of an applied magnetic field~\cite{teitel_josephson-junction_1983}. In $O(2)$ Josephson junction arrays, in the absence of an applied magnetic field, the concentrations of topological defects with equal and opposite signs are equal~\cite{gantmakher_superconductor-insulator_2010, teitel_josephson-junction_1983}. The entire concentration of topological defects binds into low-energy paired configurations below a critical temperature, via a BKT transition, and no unpaired topological defects persist to the ground state. Similarly, in the absence of geometrical frustration, crystalline ground states are achieved by a BKT-like transition and are not plagued by topological defects.

Geometrical frustration is evident in three-dimensions when the preferred orientational order of atomic clustering ($H\in SO(3)$) is incompatible with a space-filling crystal~\cite{nelson_symmetry_1984, sadoc_geometrical_2006} (e.g., Fig.~\ref{fig:frustrated_plaquette} (A)). Geometrical frustration, evident in flat space, is relieved by allowing for positive curvature to enter into each of the atomic clusters~\cite{venkataraman_beyond_1989}. This produces a tessellation of the three-dimensional space of constant positive curvature, as $\mathcal{M}=\mathbb{S}^3/H'$, that is known as a polytope~\cite{nelson_symmetry_1984} (e.g., Fig.~\ref{fig:frustrated_plaquette} (B)). In cases of geometrical frustration, owing to a curvature mismatch between the tessellation of $\mathcal{M}$ and flat space,  $\mathcal{M}$ cannot be flattened into Euclidean space without the introduction of topological defects. Because $\mathcal{M}$ is not developable in flat space, there is finite positive Gaussian curvature attributed to each atomic vertex of $\mathcal{M}=\mathbb{S}^3/H'$.

With the incorporation of geometrical frustration, the finite positive Gaussian curvature that is attributed to each geometrically frustrated atomic vertex drives an asymmetry in the plasma of disclination defects towards the concentration of disclinations that concentrates negative curvature at their core~\cite{romanov_disclinations_1983, sadoc_geometrical_2006,mosseri_geometrical_2008}. These excess negative disclinations balance out the positive curvature that is attributed to geometrically frustrated atomic vertices. Thus, in cases of geometrical frustration, the biased nature of the plasma of disclinations ensures that the three-dimensional Euclidean space remains flat on average~\cite{nelson_symmetry_1984, nelson_liquids_1983}.

The plasma of concomitant third homotopy group defects also become biased with the incorporation of geometrical frustration, which generates finite kinetic energy effects. This is akin to the shifting of the vortex-antivortex plasma in charged $O(2)$ Josephson junction arrays that express magnetic frustration in the presence of an applied magnetic field~\cite{gantmakher_superconductor-insulator_2010, teitel_josephson-junction_1983}. In the range of dominant potential energy,  in solidifying undercooled atomic liquids that are geometrically frustrated, excess negatively signed topological defects are unable to form low-energy paired configurations on crystallization and will persist to the solid ground state. In the solid state, these unpaired topological defects are pinned within a periodic arrangement so that the crystalline ground state is one of zero configurational entropy.

In three-dimensions, this periodic arrangement of topological defects (in geometrically frustrated crystalline solids) is known as the ordered major skeleton network~\cite{frank_complex_1959, frank_complex_1958} that consists of negative disclination lines~\cite{nelson_liquids_1983}. In general, geometrically frustrated crystalline structures are known as topologically close-packed (TCP) crystals which, in contrast to FCC and HCP packings, are comprised exclusively of tetrahedral interstices~\cite{de_graef_structure_2012}, i.e., without octahedral interstices. Frank-Kasper structures~\cite{frank_complex_1959, frank_complex_1958} are a particular example of TCP crystalline solid states in three-dimensions, that express preferred icosahedral local orientational order ($Y\in SO(3)$). For an example of the ordered major skeleton network of negative disclination lines in a three-dimensional Frank-Kasper structure, see Fig.~\ref{fig:curving_wedge} (A).

Figure~\ref{fig:curving_wedge} (B) depicts an anticipated phase diagram of an $O(4)$ quantum rotor model in four-dimensions, in the limit of dominant potential energy, in coordinates of temperature versus geometrical frustration. Importantly, the quantity of geometrical frustration is not a continuous variable because the preferred orientational symmetry of atomic clustering $H\in SO(3)$ is a discrete subgroup of rotations in three-dimensions. For geometrically frustrated atomic clusters, the positive Gaussian curvature attributed to each atomic vertex is inversely proportional to the radius~\cite{sethna_relieving_1983} of $\mathcal{M}=\mathbb{S}^3/H'$. That is, if fewer vertices are identified (i.e., for smaller symmetry groups $H\in SO(3)$) the radius of $\mathcal{M}$ is smaller and consequently the positive Gaussian curvature attributed to each geometrically frustrated atomic vertex increases.

With increasing geometrical frustration, the crystallization transition temperature is suppressed as the plasma of topological defects becomes increasingly biased. With the incorporation of geometrical frustration, topological defects persist to the crystalline ground state such that it is no longer one of perfect orientational order. Specifically, the set of scalar phase angle parameters that characterize the orientational order parameter will vary from site to site in order to facilitate the incorporation of an ordered major skeleton network and a periodic arrangement of third homotopy group defects. 

Ultimately, with increasing geometrical frustration, the spacing between topological defects becomes reduced until above a critical value of geometrical frustration the ground state is no longer crystalline. At zero temperature, the suppression of crystallinity at a critical value of geometrical frustration is a higher-dimensional analogue to the suppression of the superconducting ground state with critical magnetic frustration in charged $O(2)$ Josephson junction arrays (at the superconductor-to-superinsulator transition).

  \subsection{Glass Transition}
  \label{sec:glass_BKT}

In this section, it is suggested that the solidification of undercooled atomic liquids in which the atomic clusters are not strongly interacting may be described mathematically by the minimization of the kinetic energy term of relevant $O(4)$ quantum rotor models. By the minimization of the kinetic energy term, an orientationally-disordered low-temperature solid state is anticipated that is a ``dual'' to the crystalline solid state (as a consequence of the uncertainty relations~\cite{sachdev_quantum_2011} that apply to the model). Orientationally-disordered low-temperature states of four-dimensional $O(4)$ plastic crystal phases may be called ``orientational glasses,'' and corresponding three-dimensional systems are structural glasses (e.g., continuous random network glasses and metallic glasses).

Just below the melting temperature, in the range of dominant kinetic energy, internal relaxations by molecular rearrangements are necessary to remain in the undercooled state of metastable equilibrium. In these glass-forming systems, internal relaxations are thermally activated such that average structural relaxation times follow Arrhenius's law~\cite{angell_relaxation_1990}. As the temperature is lowered below some critical value (the glass transition temperature), a glass-forming undercooled system will fall out of metastable equilibrium and forms an orientationally-disordered solid state\footnote{By convention, glass formation occurs as the viscosity rises above a critical value ($10^{13}$ Poise).} that can resist shear deformations.

In the absence of potential energy effects, a maximally orientationally-disordered low-temperature solid state is anticipated in which no internal relaxations have occurred within the undercooled system prior to glass formation. Maximal orientational disorder implies that all condensed particles should become thermally pinned at the glass transition, owing to the symmetry of the uncertainty relations that apply to the $O(4)$ quantum rotor model~\cite{sachdev_quantum_2011}, in the absence of potential energy effects. This process may be considered as akin to the formation of neutral charge dipoles at the transition to the phase-incoherent superinsulating low-temperature state in charged $O(2)$ Josephson junction arrays. In this way, non-crystalline solid states may be viewed as higher-dimensional (uncharged) analogues to superinsulating ground states of Cooper pairing that can be realized for complex ordered systems in restricted dimensions~\cite{vinokur_superinsulator_2008, baturina_superinsulatorsuperconductor_2013}. 

With the incorporation of finite potential energy effects, the glass transition temperature is suppressed and so the system may become increasingly undercooled. This is possible because potential energy effects, due to coupling between neighboring atomic clusters, allow for the necessary internal relaxations to occur that ensure an adequate sampling of configurations such that the undercooled system remains ergodic~\cite{langer_mysterious_2007} to lower temperatures. In the limit that the undercooled system is driven towards the ``self-dual critical point,'' at which point the potential and kinetic energies become comparable, a hypothetical ``ideal glass'' can be achieved that is entirely internally relaxed and exhibits the maximum orientational correlations of any glass structure.

\begin{figure}[b!]
  \centering
\includegraphics[scale=.6]{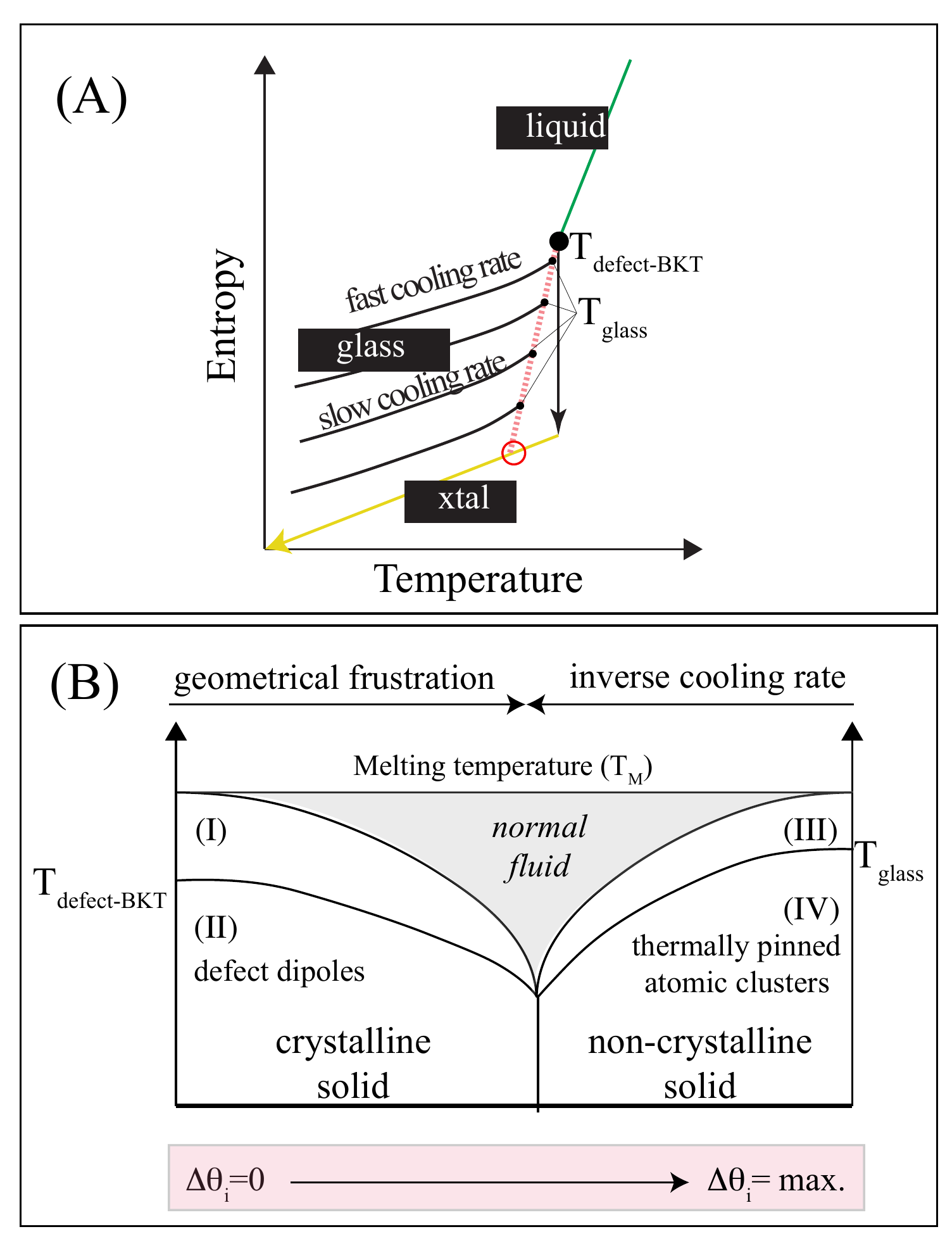}
\caption{(A) With slower cooling rates, glass-formers can become increasingly undercooled. An entropy paradox occurs in the limit of an infinitesimal cooling rate (open circle). (B) A schematic phase diagram for solidification of undercooled atomic liquids in three-dimensions, using an $O(4)$ quantum rotor model, plotted in coordinates of $T$ versus the ratio of importance of kinetic and potential energies.  }
\label{fig:full_phase_diagram}
\end{figure}

In the laboratory, the cooling rate is the most obvious tuning parameter that can be utilized to drive a glass-forming system to become increasingly undercooled (Fig.~\ref{fig:full_phase_diagram} (A)). This is because the cooling rate competes with the thermally activated processes of internal relaxation, such that the system may only undergo internal relaxation processes if enough time is allowed by the cooling rate. Eventually, as the temperature is lowered below a critical value, molecules rearrange so slowly that the undercooled system will become unable to adequately sample configurations in the time allowed by the cooling rate and it will become frozen (by falling out of ergodicity). In the hypothetical limit of an infinitely slow cooling rate, an undercooled system may become entirely internally relaxed and may form an ``ideal glass'' at the ``self-dual critical point.''

As a glass-forming system falls out of metastable equilibrium, on cooling past the glass transition temperature, condensed particles throughout the system that are not internally relaxed become thermally pinned. Thus, by the inclusion of finite potential energy effects, there is a trade-off between the thermal pinning of condensed particles at the glass transition (that are not internally relaxed) and the development of orientational correlations between neighboring atomic clusters (by internal relaxations prior to glass formation). This can be viewed as a consequence of the symmetric uncertainty relations~\cite{sachdev_quantum_2011}, between the amplitude and scalar phase angle parameters, that apply to the $O(4)$ quantum rotor model. These arguments follow as an extension of the quantum-mechanical arguments for the possibility of the existence of a superinsulating low-temperature state of charged $O(2)$ Josephson junction arrays~\cite{vinokur_superinsulator_2008, baturina_superinsulatorsuperconductor_2013}. 

A schematic of the anticipated phase diagram for three-dimensional solidification is shown in Figure~\ref{fig:full_phase_diagram} (B). As a consequence of the inability to simultaneously minimize the kinetic and potential energy terms of $O(4)$ quantum rotor model Hamiltonians, in addition to crystalline solid states (perfect and topologically close-packed), non-crystalline solid states can be realized as the temperature is lowered below a critical value. In three-dimensions, the convergence of the crystallization and glass transition temperatures at a finite temperature at the ``self-dual critical point'' corresponds to a first-order transition between crystalline and glassy solid states~\cite{baturina_superinsulatorsuperconductor_2013}. This finite temperature corresponds to the well-known Kauzmann temperature~\cite{kauzmann_nature_1948, stillinger_supercooled_1988, speedy_kauzmanns_2003} for which a paradox arises wherein the difference in configurational entropy of a non-crystalline system and its crystalline counterpart is zero.

In the range of dominant potential energy, a defect-driven Berezinskii-Kosterlitz-Thouless (BKT) transition separates an undercooled system (Region (I)) from a low-temperature orientationally-ordered solid state (Region (II)). In the absence of kinetic energy effects, a crystalline ground state is favored that is free of topological defects ($\Delta\theta_i=0$ for $i=0,1,2$). In the range of dominant potential energy, geometrical frustration skews the concentration of topological defects towards those of a particular sign and the ground state is topologically close-packed (e.g., Frank-Kasper structures). 

In the range of dominant kinetic energy, just below the melting temperature, an undercooled liquid forms in which there is little interaction between atomic clusters (Region (III)). The glass transition temperature, which marks the freezing of this system into a particular non-equilibrium configuration (Region (IV)), is sensitive to the cooling rate. A maximally orientationally-disordered solid state ($\Delta\theta_i=\text{max.}$ for $i=0,1,2$) is obtained in the absence of interactions between atomic clusters in cases of arbitrarily fast cooling rates. In contrast, non-crystalline solid states that form in the hypothetical limit of an infinitely slow cooling rate (i.e., an ``ideal glass'' that forms at the ``self-dual critical point'') are entirely internally relaxed yet lack translational order. In physical systems, this entropy paradox at the ``self-dual critical point''  is always avoided because any real glass transition occurs at a temperature above the Kauzmann temperature which would require a cooling rate longer than the duration of the universe to achieve.

 \begin{figure}[t!]
  \centering
\includegraphics[scale=.9]{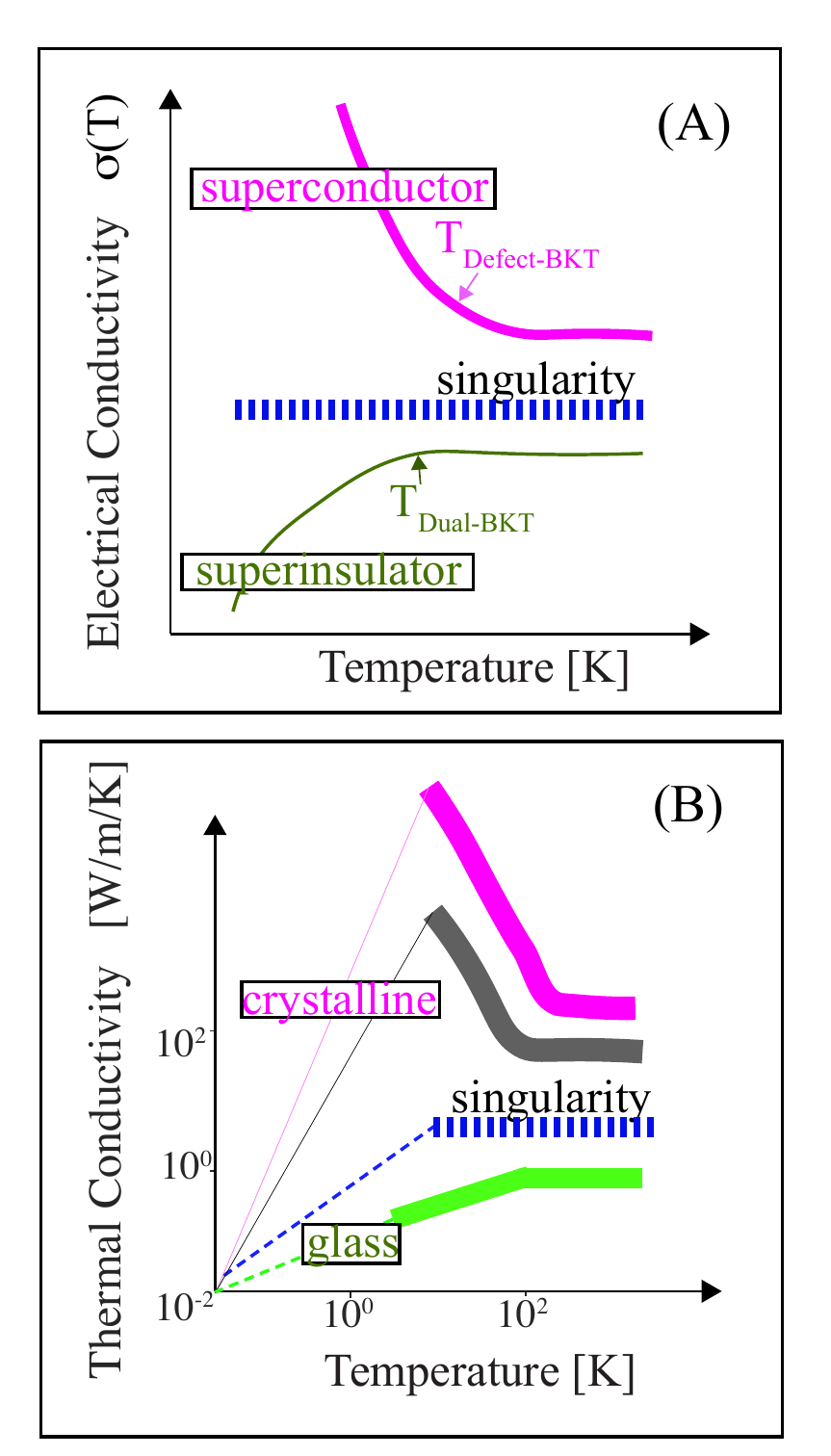}
\caption{(A) The superconductor-to-superinsulator transition in $O(2)$ Josephson junction arrays is displayed in the temperature dependence of the electrical conductivity. A superconductor forms in the range of dominant potential energy, and a superinsulator, which exhibits ``zero electrical resistance'' is preferred in the range of dominant kinetic energy. (B) Anticipated thermal conductivity ($\kappa$), across a transition between crystalline and non-crystalline ground states, above approximately 50 K. With the incorporation of geometrical frustration into the crystalline ground state, the overall magnitude of $\kappa$ should become suppressed. In contrast to crystalline solids, $\kappa$ of non-crystalline solids decreases with decreasing temperatures. The magnitude of $\kappa$ of non-crystalline solids is anticipated to be proportional to the degree of internal relaxation prior to glass formation.  }
\label{fig:thermal_prop}
\end{figure}

\section{Emergent thermal transport properties}

The thermal properties of solid state systems are intimately related to the structure that forms from the undercooled atomic liquid. Applying the topological framework developed herein, the inverse temperature dependence of the thermal conductivity of crystalline and non-crystalline solid states (above approximately 50 K) may be viewed as a consequence of the realization of distinct low-temperature states of $O(4)$ quantum rotor models. It follows that the thermal transport properties of solid states may be viewed as analogous to the well-studied electrical transport properties of charged $O(2)$ Josephson junction arrays~\cite{poran_quantum_2017}, which display a singularity at the superconductor-to-superinsulator transition~\cite{vinokur_superinsulator_2008} (Fig.~\ref{fig:thermal_prop} (A)). Likewise, a singularity in the thermal transport properties as a function of temperature is anticipated at the ``self-dual critical point'' that belongs to $O(4)$ quantum rotor models.  

Anticipated thermal conductivities of crystalline and non-crystalline solid states, above approximately 50 K, are shown in Fig.~\ref{fig:thermal_prop} (B). In the absence of geometrical frustration, a perfect crystalline ground state that is free of topological defects is achieved for which the thermal conductivity rises to a maximum value (pink). With the incorporation of geometrical frustration, the presence of frustration induced topological defects generates a finite uncertainty in the global orientational order parameter (despite their periodic arrangement in the crystalline ground state); this leads to the suppression of the overall solid state thermal conductivity (grey).

With a critical value of geometrical frustration, the ground state is no longer crystalline and characteristic features of translational order are no longer displayed in the thermal transport properties; this leads an anticipated a singularity in the thermal transport properties at the at the ``self-dual critical point'' (dashed). Despite lacking translational order, this system does express orientational order and therefore acts as a bridge between the crystalline and non-crystalline structures. This hypothetical non-crystalline solid state (i.e., ``ideal glass''), should present thermal transport properties that are similar to quasicrystals (which have orientational order but not translational order). The thermal transport properties of quasicrystalline materials have been shown to be comparable to those observed in amorphous systems~\cite{michalski_comparison_1989, michalski_thermal_1992}.

In contrast to crystalline solids, collective excitations (phonons) are unable to exist in non-crystalline solids. Above approximately 50 K, the thermal conductivity of non-crystalline solids decreases with decreasing temperatures and is well-described by Einstein's picture of a random walk of thermal energy between localized oscillators vibrating with random phases~\cite{cahill_lattice_1988, cahill_lower_1992}. Within the context of the topological framework presented herein, this random phase assumption has its origin in the fact that the kinetic energy term in the $O(4)$ Hamiltonian favors orientational disorder in the low-temperature solid state. In the absence of potential energy effects, the solid that forms is a physical realization of a system Einstein oscillators are entirely uncoupled and thus cannot transmit thermal energy throughout the system. Real non-crystalline solids (green) exhibit some orientational correlations between neighboring atomic clusters, that allow for heat transfer by a random walk mechanism.

\section{Summary and conclusions}

The current research is based on seminal work from across a wide range of fields, in order to offer a perspective on the topological origins of the development of the solid state and its transport properties. In particular, we have:
\begin{itemize}
\item introduced a quaternion orientational order parameter for solidification, which characterizes the topological structure of undercooled atomic fluids.
\item identified four- and three-dimensions as restricted dimensions for quaternion ordered systems, such that O(4) quantum rotor models apply. 
\item identified an important role of third homotopy group topological defects in solidification. 
\item utilized uncertainty relations that apply to quaternion ordered systems, between amplitude and scalar phase angle parameters, to invoke a duality between crystallization and the glass transition that can be realized in restricted dimensions. This has enabled the development of a phase diagram for solidification, in analogy to the phase diagram for Josephson junction arrays in the vicinity of the superconductor-to-superinsulator transition.
\item classified ordered major skeleton networks, induced by geometrical frustration, as a consequence of finite kinetic energy effects in $O(4)$ quantum rotor models.
\item introduced a topological perspective of the well-known Kauzmann point, relating it to the ``self-dual critical point'' that belongs to $O(4)$ quantum rotor models.
\item compared the inverse thermal transport properties of crystalline and non-crystalline solid states, above approximately 50 K, to the inverse temperature dependence of the electrical transport properties of Josephson junction arrays across the superconductor-to-superinsulator transition.
\end{itemize}

\noindent The topological framework presented in this work has enabled the generalization of the notion of lower critical restricted dimension among: real, complex, and quaternion algebra domains (see Fig.~\ref{fig:phase_trans}). In this way, undercooling below the melting temperature in three-dimensions may be viewed as a topological consequence due to the dimensionality of the orientational order parameter.

\section{Acknowledgement}
The authors would like to acknowledge Michael Widom for useful discussions on this research topic. C.S.G thanks the NASA's Office of Graduate Research, Space Technology and Research Fellowship (NSTRF), for funding.


\bibliography{\jobname}


\end{document}